\documentclass[runningheads]{llncs}

\usepackage[T1]{fontenc}
\usepackage{graphicx}
\usepackage[firstpage]{draftwatermark}
\usepackage{xcolor}
\usepackage{color, colortbl}
\usepackage{bbding}

\usepackage{amssymb}
\usepackage{amsmath}

\usepackage[hidelinks]{hyperref}
\usepackage{cleveref}

\usepackage{tikz}
\usepackage{pgfplots}
\usepackage{subcaption}
\usetikzlibrary{shapes.geometric, arrows,calc,fit,backgrounds}

\usepackage{wrapfig,lipsum,booktabs}

\usepackage{xspace}

\usepackage{pifont}
\newcommand{\cmark}{\ding{51}}
\newcommand{\xmark}{\ding{55}}
\newcommand{\bmark}{$\bullet$}

\usepackage{fontawesome5}

\usepackage{syntax}
\usepackage{listings}

\newcommand{\issy}{\textsf{Issy}\xspace}
\newcommand{\llissy}{\textsf{LLissy}\xspace}

\newcommand{\rpgstela}{\textsf{rpg-STeLA}\xspace}
\newcommand{\rpgsolve}{\textsf{rpgsolve}\xspace}
\newcommand{\tslmtrpg}{\textsf{tslmt2rpg}\xspace}

\newcommand{\muval}{\textsf{MuVal}\xspace}
\newcommand{\simsynth}{\textsc{SimSynth}\xspace}
\newcommand{\temos}{\textsf{temos}\xspace}
\newcommand{\raboniel}{\textsf{Raboniel}\xspace}
\newcommand{\gensys}{\textsc{GenSys}\xspace}
\newcommand{\gensysltl}{\textsc{GenSys-LTL}\xspace}
\newcommand{\cesars}{tools in~\cite{RodriguezS23,RodriguezS24,RodriguezGS24}}

\newcommand{\rpltl}{{RP{-}LTL}}

\SetWatermarkAngle{0}
\SetWatermarkText{\raisebox{17.6cm}{
\href{https://doi.org/10.5281/zenodo.15308725}{\includegraphics{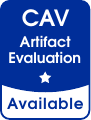}}
\hspace{8cm}
\includegraphics{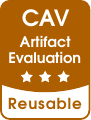}
}}

\begin{document}

\title{Issy: A Comprehensive Tool for\\Specification and Synthesis of\\Infinite-State Reactive Systems}
\titlerunning{Issy: Specification and Synthesis of Infinite-State Reactive Systems}

\author{
Philippe Heim\Envelope\orcidID{0000-0002-5433-8133} \and
Rayna Dimitrova\orcidID{0009-0006-2494-8690}
}

\authorrunning{P. Heim \and R. Dimitrova}

\institute{
CISPA Helmholtz Center for Information Security, Saarbr\"ucken, Germany
\email{\{philippe.heim, dimitrova\}@cispa.de}
}

\maketitle

\begin{abstract}
The synthesis of infinite-state reactive systems from temporal logic specifications or infinite-state games has attracted significant attention in recent years, leading to the emergence of novel solving techniques. 
Most approaches are accompanied by an implementation showcasing their viability on an increasingly larger collection of benchmarks.
Those implementations are --often simple-- prototypes. 
Furthermore, differences in specification formalisms and formats make comparisons difficult, and writing specifications is a tedious and error-prone task.

To address this,  we present \issy{},  a tool for specification, realizability, and synthesis of infinite-state reactive systems.  
\issy{} comes with an expressive specification language that allows for combining infinite-state games and temporal formulas,  thus encompassing the current formalisms.
The realizability checking and synthesis methods implemented in \issy build upon recently developed approaches and extend them with newly engineered efficient techniques, offering a portfolio of solving algorithms.
We evaluate \issy on an extensive set of benchmarks,  demonstrating its competitiveness with the state of the art.
Furthermore,  \issy{} provides tooling for a general high-level format designed to make specification easier for users.  It also includes a compiler to a more machine-readable format that other tool developers can easily use, which we hope will lead to a broader adoption and advances in infinite-state reactive synthesis.
\end{abstract}

\section{Introduction}\label{sec:intro}
Reactive systems are computational systems that constantly interact with their environment and run indefinitely.  Notable examples include communication protocols and controllers for embedded systems or robots.  Reactive synthesis is the problem of automatically generating correct-by-construction reactive systems from formal specifications describing the system's desired behavior.
Many target applications of synthesis require the treatment of \emph{infinite-state} models,  as they operate with unbounded data such as integers.
For this reason, the synthesis of reactive systems over infinite data domains has received increasing  attention over the last years.  In this paper, we present \issy, a comprehensive open-source  tool for the specification and synthesis of infinite-state reactive systems,  that builds upon recently developed approaches and newly engineered efficient techniques.
\issy{} comes with an expressive specification language that encompasses current formalisms,  thus providing a basis for the further development of synthesis tools.

\begin{table}[t!]\footnotesize
\begin{center}
\begin{tabular}{|p{2.6cm} || p{2.6cm} ||p{.43cm} | p{.3cm} | p{.6cm}  || 
p{.5cm}| p{.55cm} ||
 p{.23cm} | p{.23cm} | p{.23cm} || p{2.25cm} |} 
\hline 
 Tool & Specification Type & 
 SR & B & LTL & 
 Inf Inp& Inf Out & $\mathbb B$ & $\mathbb Z$ & $\mathbb R$ &
 Input Format \\
 \hline\hline
 \rowcolor{blue!10}
  \issy & Combined games \& \rpltl\ & 
 \cmark & \cmark & \cmark &  
 \cmark & \cmark & \cmark & \cmark & \cmark &
  \issy,  \llissy, RPG, TSL-MT\\  
 \hline\hline
 \rpgsolve~\cite{HeimD24} & RPG  & 
 \cmark & \cmark & \xmark & 
 \cmark & \xmark & \cmark & \cmark & \cmark & 
 RPG \\
 \hline
 \rpgstela~\cite{SchmuckHDN24} & RPG & 
 \cmark &\cmark & \xmark  & 
 \cmark & \xmark & \cmark & \cmark & \cmark &
 RPG \\
 \hline
 \tslmtrpg~\cite{HeimD25}~+~\cite{HeimD24} & TSL-MT&
 \cmark & \cmark & \cmark &  
 \cmark & \xmark & \cmark & \cmark & \cmark & 
 TSL-MT var. \\
 \hline\hline
  sweap~\cite{AzzopardiPSS24} & Programs + LTL & 
  \cmark & \cmark & \cmark & 
  \xmark & \xmark & \cmark & \cmark & \xmark &
  custom format \\
  \hline 
 \raboniel~\cite{MaderbacherB22} & TSL-MT &
 \cmark &\cmark & \cmark  & 
 \cmark & \xmark & \xmark & \cmark & \cmark & 
 TSL-MT var. \\
 \hline
\temos~\cite{ChoiFPS22} & TSL-MT& 
 \cmark & \cmark & \cmark & 
 \cmark & \xmark & \cmark & \cmark & \xmark & 
 TSL-MT  var. \\ 
 \hline\hline
  \gensys~\cite{SamuelDK21} & Games & 
  \cmark & \xmark & \xmark & 
  \cmark & \cmark & \cmark & \cmark & \cmark &
  --- \\
 \hline
 \gensysltl~\cite{SamuelDK23} & Games &
 \cmark & \cmark & \cmark  & 
 \cmark & \cmark & \cmark & \cmark & \cmark &
 --- \\
 \hline
 gr1mT~\cite{MaderbacherWB24} & GR(1) + data &
 \cmark & \cmark & \xmark  & 
 \cmark & \xmark & \cmark & \cmark & \cmark & 
 unkown \\
 \hline\hline
 \cesars & LTL$_{\mathcal T}$  &
 \cmark & \cmark & \cmark &
 \cmark & \cmark & \cmark & \cmark  & \cmark & 
 unkown \\
 \hline\hline
 \muval~\cite{UnnoTGK23} & $\mu$CLP formulas &
 \cmark & \cmark & \cmark  & 
 \cmark &\cmark & \xmark & \cmark & \cmark & 
 custom format\\
 \hline
 \simsynth~\cite{FarzanK18} & linear arith.  games &
 \cmark & \xmark & \xmark  & 
 \cmark & \cmark & \cmark & \cmark & \cmark & custom format\\
\hline
\end{tabular}
\end{center}
\caption{Supported specification type (temporal logic or games) and objectives (Safety/Reachability,  Deterministic B\"uchi,  LTL),  support for infinite domains of system input and output, supported data types ($\mathbb B, \mathbb Z, \mathbb R$), input format.}
\label{table:compare-input}
\end{table}

Requirements for reactive systems are typically specified using \emph{temporal logics},  such as Linear Temporal Logic (LTL)~\cite{Pnueli77} in the case of finite-state systems. 
 Alternatively,  the synthesis problem can be described as a \emph{two-player game} modelling the interaction between a system and its environment.  These specification formalisms have been extended to the setting of infinite-state systems, resulting in temporal logics such as TSL-MT~\cite{FinkbeinerHP22}, LTL$_\mathcal T$~\cite{RodriguezS23}, and \rpltl~\cite{HeimD25},  on the one hand,  and infinite-state game models, such as reactive program games (RPGs)~\cite{HeimD24} on the other.
\Cref{table:compare-input} summarizes the types of specifications used in the main existing  prototype tools for realizability and synthesis of infinite-state reactive systems,  and that of our new tool \issy presented in this paper.
Most of the tools fall into one of two categories: those that support temporal logic formulas (\temos~\cite{ChoiFPS22},  \raboniel~\cite{MaderbacherB22}, the \cesars, sweap~\cite{AzzopardiPSS24}, \tslmtrpg~\cite{HeimD25}) and those using directly two-player games
(\gensys~\cite{SamuelDK21}, \gensysltl~\cite{SamuelDK23}, gr1mT~\cite{MaderbacherWB24}, \simsynth~\cite{FarzanK18},  \rpgsolve~\cite{HeimD24},  \rpgstela~\cite{SchmuckHDN24}).
However, different types of requirements are more naturally modelled in one formalism or the other. 
For example,  constraints that depend heavily of the systems' state or execution phase (such as, for instance, the available moves of a robot) are often difficult to express in temporal logic, and result in long and complex formulas.  High-level mission requirements, on the other hand,  are more naturally formalized in temporal logics.  Motivated by this, we developed \issy with support for a new input format that unites both specification paradigms.  
Often, even tools using the same specification logic have different input formats, such as for example the tools in \Cref{table:compare-input} using TSL-MT. 
In contrast to the case for finite-state systems, where an established specification format, TLSF~\cite{JacobsPS23},  exists and is used in SYNTCOMP~\cite{syntcomp},  there is no such common format for  infinite-state reactive systems. 
We envision that the \issy framework is a major step towards filling this gap.
The \issy input format  supports the main types of synthesis objectives,  possibly infinite domains for both the input and the output variables of the specified system,  and three basic data types (bool, int, and real).  As \Cref{table:compare-input} shows,  the specification capabilities of \issy strictly subsume those of the  existing tools.

\begin{table}[t!]\footnotesize
\begin{center}
\begin{tabular}{| p{2.725cm} || p{2.72cm} | p{.72cm} ||  p{1.52cm} ||  p{.725cm} | p{.7cm} | p{.927cm} || p{.927cm}|} 
\hline 
 Tool & Technique & Unb. Loop& Synthesis &LTL Synt. & SMT  & LTL to Aut & Open-Source \\
  \hline\hline
  \rowcolor{blue!10}
   \issy & acceleration-based f.p.~computation & \cmark & \cmark (C code)& & \bmark   & \bmark & \cmark \\
    \hline\hline
  \rpgsolve~\cite{HeimD24} & acceleration-based f.p.~computation  & \cmark &\cmark & & \bmark & & \cmark \\
 \hline
  \rpgstela~\cite{SchmuckHDN24} & \cite{HeimD24}  + abstraction & \cmark & \xmark& & \bmark&  & \cmark \\
 \hline
  \tslmtrpg\cite{HeimD25}\newline  + \rpgsolve & monitor-enhanced symb. game constr. & \cmark & \cmark & & \bmark & \bmark& \cmark \\
   \hline\hline
  sweap~\cite{AzzopardiPSS24} & abstraction to LTL  & \cmark &\cmark & \bmark & \bmark  & & \cmark \\
 \hline
  \raboniel~\cite{MaderbacherB22} & abstraction to LTL & \xmark & \cmark (Python)& \bmark & \bmark   & & \cmark\\
 \hline
 \temos~\cite{ChoiFPS22} & abstraction to LTL  & \xmark &\cmark (several)& \bmark & \bmark   & & \cmark\\
 \hline\hline
  \gensys~\cite{SamuelDK21} & naive f.p.~comp.& \xmark & \cmark & & \bmark   & & \cmark\\
 \hline
 \gensysltl~\cite{SamuelDK23} & naive f.p.~comp. & \xmark & \cmark & & \bmark   & \bmark & \cmark\\
 \hline
  gr1mT~\cite{MaderbacherWB24} & GR(1) f.p.~comp. & \xmark & \cmark & & \bmark & & \xmark \\
 \hline\hline
  tool in~\cite{RodriguezS23} & abstraction to LTL  & --- &  \xmark & \bmark & \bmark   & & \xmark\\\hline
  tool in~\cite{RodriguezS24} & abstraction to LTL  & --- &  \cmark & \bmark & \bmark   & & \xmark\\\hline
  tool in~\cite{RodriguezGS24} & abstraction to LTL + Skolem fun. syn. & --- &  \cmark (C code) & \bmark & \bmark   & & \xmark\\
 \hline\hline
  \muval~\cite{UnnoTGK23} & constraint solving & \cmark &\xmark & & \bmark   & & \cmark\\
 \hline
  \simsynth~\cite{FarzanK18} & constraint solving & --- &\cmark & & \bmark  & & \cmark\\
\hline
\end{tabular}
\end{center}
\caption{Comparison of main techniques, capabilities,  technologies, availability}
\label{table:compare-techniques}
\end{table}

The synthesis problem for infinite-state systems is in general undecidable.  
From \Cref{table:compare-input},  only~\cite{RodriguezS23,RodriguezS24,RodriguezGS24}  considers a decidable restriction of the problem. The others implement different incomplete techniques summarized in \Cref{table:compare-techniques}.
One of the common approaches,  used in \temos, \raboniel,  and sweap,  is abstraction to synthesis from LTL specifications, accompanied by some form of specification refinement.  Alternatives include  fixpoint-based game-solving as~\cite{SamuelDK21,SamuelDK23,MaderbacherWB24}, and constraint solving as in \simsynth and \muval, the last of which is a tool for solving first-order fixpoint constraints.
In~\cite{HeimD24} we proposed a technique for solving infinite-state games that aims to address one of the limitations of prior abstraction and fixpoint-based approaches, namely, that they usually diverge on game-solving tasks that require reasoning about the unbounded iteration of strategic decisions.  
The core of~\cite{HeimD24} is a technique called \emph{attractor acceleration} that employs \emph{ranking arguments} to improve the convergence of symbolic game-solving procedures. 
\cite{AzzopardiPSS24}  also addresses unbounded behavior,  in the context of abstraction-based methods by introducing the so called liveness refinement. 
Column ``Unb.  Loop'' indicates which of the techniques handle unbounded strategy loops.  Further, the table indicates whether the tool performs synthesis (or only checks realizability,  that is, the existence of an implementation for the specification).  We also indicate the main technologies (LTL synthesis,  SMT, translation of LTL to automata) used by each tool, and whether the tool is available open-source. 

 \issy builds on the acceleration technique in~\cite{HeimD24},  but in addition to the new input format,  integrates methods and ideas from our recent work~\cite{SchmuckHDN24} and \cite{HeimD25},  as well as novel techniques discussed in \Cref{sec:solver}. 
We evaluate \issy on an extensive set of benchmarks,  demonstrating its competitiveness with the state of the art.

\section{The Issy Format}\label{sec:format-issy}
The \issy input format has the key advantage that it combines two modes for specification of synthesis problems for infinite-state reactive systems: temporal logic formulas,  and two-player games, both over variables with infinite domains, such as integers or reals.  
The advantages of this mutli-paradigm specification format are two-fold.
First, it allows specification designers to specify requirements in a less cumbersome way.  
For example,  constraints that  depend on the system's state, or encode behaviour in different phases,  are usually easier to specify as games.  On the other hand, mission specifications such that under certain assumptions the system must eventually stabilize, or that some tasks should be carried out repeatedly, are often more concisely expressed in temporal logic.

\definecolor{codegray}{rgb}{0.5,0.5,0.5}
\definecolor{codepurple}{rgb}{0.58,0,0.82}
\definecolor{backcolour}{rgb}{0.95,0.95,0.92}

\lstdefinestyle{mystyle}{
    backgroundcolor=\color{backcolour},   
    commentstyle=\color{codegray},
    keywordstyle=\color{blue},
    numberstyle=\tiny\color{codepurple},
    basicstyle=\linespread{0.9}\ttfamily\footnotesize,
    breakatwhitespace=false,         
    breaklines=true,                 
    captionpos=b,                    
    keepspaces=true,                 
    numbers=left,                    
    numbersep=5pt,                  
    showspaces=false,                
    showstringspaces=false,
    showtabs=false,                  
    tabsize=1,
    morecomment=[l]{//},
morecomment=[s]{/*}{*/},
basewidth = {.47em}
}

\lstset{style=mystyle,  morekeywords={game,  input, state, formula, def, assert, assume, keep, from,to, with, Safety,loc}}
\begin{center}
\begin{lstlisting}[label=lst:example,caption=Example specification in \issy format.,escapeinside={@}{@}]
input real add  // Real-valued input variables
input real rem  // Input variables are global for all formulas and games.
                // The values of input variables are picked by the environment
                // at every step and they are not stored as part of the state.
state real load1  // Real-valued state variables
state real load2  // State variables are global for all formulas and games.
state real rem1   // They are controlled by the system, choosing the next 
state real rem2   // values based on the current state and environment input. 

/* Specifications consist of formulas and game specifications blocks. Those 
   blocks are interpreted conjunctively. A single formula is an implication 
   between  conjuncted assumptions and conjuncted assertions (guarantees). */
formula {
 /* Assumption: From some time point on, the environment will always set the input variable add to be less than or equal to zero. */
 assume F G [add <= 0]
 /* Guarantee: From some point on, load1 and load2 will always be zero. */     
 @\label{lst:lassert}@assert F G ([load1 = 0] && [load2 = 0])
}

// Macros to make the specification easier to read
@\label{lst:baldef}@def balanced  = [load1 >= load2] && [load1 <= 2 * load2] 
              ||[load2 >= load1] && [load2 <= 2 * load1]
def addtoone  = [load1' = load1 + add] && [load2' = load2] 
              ||[load2' = load2 + add] && [load1' = load1]
def validrem  = [rem >= 0.1] && [rem <= load1 + 2/3 * load2]
def decrease  = [load1' = load1 - rem1'] && [rem1' + rem2' = rem]
              &&[load2' = load2 - 3/2 * rem2'] 

/* Two-player game with locations init, lbal, lrem, done and err, and safety winning condition for the system, requiring that err is never reached. */
game Safety from init {
  loc init 1 // When defining locations, the type of the location w.r.t. the
  loc lbal 1 // accepting condition is specified. Here, 1 means that those
  loc lrem 1 // locations are safe. The scope of each location is the
  loc done 1 // respective game. Different formulas and games are related
             // via the variables, making their combination less error-prone.
  loc err 0  // The location err is the only unsafe location in this game.

  /* The following define the possible moves in the game via pairs of locations and their transition constraints over the current state and input variables as well as the next-state variables. A move in such a game works as follows. It starts in some location and assignment to the state variables. First, the environment chooses values for the input variables. Then, the system chooses the next state values and the next location such that the respective transition constraint is satisfied. */
  from init to done with [load1 < 0] || [load2 < 0]
  from init to lbal with [load1 >= 0] && [load2 >= 0] && keep(load1 load2)
  // Conditions like the next one are not possible in TSL-MT.
  @\label{lst:lbalrem}@from lbal to lrem with [load1' + load2' = load1 + load2] 
  @\label{lst:lremerr}@from lrem to err  with !balanced 
  from lrem to done with balanced &&(!validrem ||([load1 = 0] && [load2 = 0]))
  @\label{lst:lrembal}@from lrem to lbal with balanced && [add > 0]  && addtoone

  from lrem to lrem with balanced && [add <= 0] && validrem && decrease
  from done to done with true 
  from err  to err  with keep(load1 load2)
}
\end{lstlisting}
\end{center}

Each of the two modes of specification can potentially offer opportunities for optimization of the synthesis tools processing these specifications. 
In~\cite{HeimD25}, we showed how the translation from \rpltl{} formulas  to  games can benefit from the high-level information present in the formula in order to simplify the  game.

\begin{figure}[t!]
\begin{center}
\input{issy-grammar-part.tex}
\end{center}
\caption{An excerpt from the \issy  format.  The full description is in~\Cref{sec:format-issy-full}.}
\label{fig:issy-grammar-part}
\end{figure}

Now, we turn to an example that illustrates and motivates the main features of the \issy format.
An excerpt of the format's grammar is given in \Cref{fig:issy-grammar-part}.

\begin{example}\label{ex:simple-example}
Consider a reactive system that has to balance the loads, \texttt{load1} and \texttt{load2}, of two components.
At any point,  the environment can increase the total load,  via the environment-controlled input variable \texttt{add}.  When that happens,  the system has to re-balance the total load by appropriate partitioning. When the load does not increase,  the system has to control the throughput of each component,  state variables \texttt{rem1} and \texttt{rem2} respectively,  in accordance with the components' speeds and the total available throughput,  \texttt{rem} controlled by the environment.
The specification of this system is given in Listing~\ref{lst:example}, and consists of variable declarations,  a formula specification,   macro definitions for better readability, and the second part of the specification given as a two-player game.

Variable declarations specify whether the variable is \texttt{\textcolor{blue}{input}} controlled by the environment, or is a \texttt{\textcolor{blue}{state}} variable controlled by the system.
The currently supported data types are \texttt{bool, int} and \texttt{real}.
The domains of variables can be further constrained in the \texttt{\textcolor{blue}{game}} specifications by additional constraints.

The \texttt{\textcolor{blue}{formula}} specification is a list of \rpltl{} formulas,  prefixed by the keywords \texttt{\textcolor{blue}{assume}} and \texttt{\textcolor{blue}{assert}},  denoting constraints on the environment and system respectively.
They use temporal operators like LTL, but with quantifier-free first-order atoms 
instead of Boolean propositions.
The assumption \texttt{F G [add <= 0]} uses the temporal operators \texttt{F} (eventually) and  \texttt{G} (globally) to state that from some point on, no more load will be added by the environment.
The assert statement in line \ref{lst:lassert} requires the system to ensure, under the above assumption, that both loads eventually stabilize at zero. 
The semantics of a \texttt{\textcolor{blue}{formula}} specification is that the conjunction of the assumptions implies the conjunction of the asserts. 

The possible actions of the system and the requirement to balance \texttt{load1} and \texttt{load2} are described by the 
\texttt{\textcolor{blue}{game}} specification in Listing~\ref{lst:example}. The game has locations \texttt{init}, \texttt{lbal}, \texttt{lrem}, \texttt{done}, \texttt{err} that are local to the game, unlike variables that are global to the whole specification.
The transitions between locations in the game are defined via  quantifier-free formulas over input, state, and next-state variables (such as \texttt{load1'}).  Nondeterminism is under the control of the system.  The \issy format enables the use of macros to improve formula readability. 
For example,  the transition in line \ref{lst:lremerr} requires the system to transition from location  \texttt{lrem} to the unsafe location \texttt{err} if the condition \texttt{balanced} defined by the macro in line \ref{lst:baldef} is violated. 
The game has a \emph{safety} winning condition,  indicated by the keyword \texttt{\textcolor{blue}{Safety}}, and defined by the natural numbers with which the locations are labelled ($0$ indicates that \texttt{err} is unsafe, while all  labelled $1$  are safe).
\end{example}

A specification can contain multiple \texttt{\textcolor{blue}{formula}} and \texttt{\textcolor{blue}{game}} components,  interpreted conjunctively.  The semantics is a two-player game defined as the product of the games for the individual formulas and all game specifications.  
\issy requires and checks that at most one of these games has a non-safety winning condition. 

The \issy specification in \Cref{ex:simple-example} shows the modelling flexibility of the format. Expressing the same requirements purely in \rpltl\ or as an RPG results in a difficult to write and understand specification,  making the specification process error-prone.  We believe that \issy  alleviates this problem to some extent, offering modularity and syntactic sugar constructs, and,  most importantly, unifying the temporal logic and game formats for 
infinite-state reactive systems.

\paragraph{The \issy compiler and the \llissy format.}
The \issy compiler,  part of our synthesis framework,  compiles specifications in  \issy format to a low-level intermediate format called \llissy,  given in \Cref{sec:format-llissy}.
The compiler checks compliance with the syntax and gives informative error messages.
The \llissy format is easier to parse, while retaining the ability to specify both logical formulas and games. 
We envision that the development of tools  for translation from various high-level specification formats to the \llissy format will enable the seamless exchange of benchmarks and experimental comparison between different tools.
\issy also accepts input directly in \issy format, as well as the older formats  \textsf{tslmt} and \textsf{rpg}.

\section{From Temporal Formulas to Games}\label{sec:to-game}
To check the realizability of specifications and synthesize reactive programs, \issy follows the classical approach of reducing the task to solving a two-player game.
To this end, it translates the specification into a symbolic synthesis game by first translating the temporal logic formulas to games, and then building their product with the rest of the specification. 
The construction of games from the  formulas follows~\cite{HeimD25}  and provides the option to build and use a  \emph{monitor} to prune/simplify the constructed game by performing first-order and temporal  reasoning during game construction. 
More concretely, a given formula is first translated to a deterministic $\omega$-automaton using \texttt{Spot}~\cite{Duret-LutzRCRAS22}.  
Then, monitors are constructed \emph{on-the-fly}, building the product between the game obtained from the automaton and the monitor. 
The product with the monitor enhances the game with semantic information~\cite{HeimD25}, resulting in the so-called \emph{enhanced game}, which is potentially easier to solve. 
As sometimes the monitor construction causes overhead, \issy has a parameter \texttt{\textcolor{blue}{-{}-pruning}} controlling its complexity,  ranging from no monitor construction (level 0), to applying powerful deduction during its construction (level 3).

The prototype \tslmtrpg~\cite{HeimD25} is restricted to the logic TSL-MT and constructs RPGs. In contrast,  the translation in \issy applies to the more general logic \rpltl, and constructs a more general form of symbolic games.  
In TSL-MT and RPGs, the system controls the state variables via a \emph{fixed finite set of possible updates}, a restriction not present in \rpltl{} and the respective symbolic games. 
For example, assertions like \texttt{x' > x} are not in the syntax of TSL-MT, and specifying the same behavior with updates would require a (possibly uncountable) infinite number of them. 
Hence, \rpltl{} lifts the imbalance in TSL-MT that the environment can pick any value (from possibly infinitely many) for the inputs, but the system can only choose from a finite set of updates.
Note that in \rpltl{}, only state variables appear primed, not environment-controlled input variables. 
If a property needs to relate input values over time,  input values need to be stored in state variables.
Hence, inputs are not part of the state unless stored explicitly, which we believe results in an intuitive notion of state.

\section{An Acceleration-Based Solver for Infinite-State Games}\label{sec:solver}
\begin{figure}[t!]
\tikzstyle{interface} = [rectangle, rounded corners, minimum width=2cm, minimum height=.6cm,text centered, draw=black, fill=red!20]
\tikzstyle{io} = [trapezium, trapezium left angle=85, trapezium right angle=95, minimum width=.5cm, minimum height=.6cm, text centered, draw=black, fill=gray!20]
\tikzstyle{process} = [rectangle, minimum width=1.5cm, minimum height=.6cm, text centered, draw=blue, fill=blue!20]
\tikzstyle{ext} = [rectangle, minimum width=1.5cm, minimum height=.6cm, text centered, draw=black]
\tikzstyle{arrow} = [thick,->,>=stealth]
\tikzstyle{doublearrow} = [thick,<->,>=stealth]

\begin{center}

\scalebox{0.95}{%
\begin{tikzpicture}[scale=0.75, node distance=2cm]
\node (issycomp) [process, text width=1.5cm] {\issy compiler};
\node (llissyprob) [io, text width=1.65cm, below of = issycomp, yshift=.7cm] {\phantom{aa} \llissy \newline specification};
\node (llissypars) [process, text width=1.5cm, below of = llissyprob, yshift=.7cm] {\llissy parser};

\node (spec) [io, text width=3.6cm, right of = llissypars, xshift=1.6cm] {Specification: Formula$_1,\ldots,$ Formula$_m$ Game$_1,\ldots,$ Game$_n$};
\node (gameprod) [process, text width=2cm, right of = spec, xshift=3cm] {Game product construction};
\node (enhgame) [io, text width=2.25cm, above of = gameprod, yshift=-.7cm, xshift=-.5cm] {Enhanced game};
\node (otfprod) [process, text width=2cm, above of = enhgame, yshift=-.5cm,xshift=.5cm,yshift=-.2cm] {On-the-fly product construction};

\node (formgame) [io, text width=.8cm,  above left of = otfprod, xshift=-1cm,yshift=-.6cm] {Game};
\node (monitor) [io, text width=1cm,  below left of = otfprod, xshift=-1cm, yshift=1.3cm] {Monitor};

\node (directconstr) [process, text width=2cm, left of = formgame, xshift=-1.5cm] {Formulas to games};
\node (monconstr) [process, text width=2cm, left of = monitor, xshift=-.5cm, yshift=0cm] {Formulas to monitors};

\node (game) [io, text width=1.5cm, below of = gameprod, yshift=.6cm, xshift=0cm] {Symbolic game};

\node (gameinterface) [interface, text width=4cm, left of = game, xshift=-1.5cm] {Game-solving interface};

\node (rpg) [io, minimum width=.5cm, left of = gameinterface, xshift=-1.05cm] {RPG};
\node (rpgpars) [ext, text width=1.5cm, above left of = rpg, xshift=-.5cm, yshift=-1cm] {RPG parser};
\node (tslmttorpg) [ext, text width=1.5cm, below left of = rpg, xshift=-.5cm, yshift=.85cm] {\tslmtrpg};

\node (solver) [process, text width=1.5cm, below of = gameinterface,  yshift=.4cm] {Game solver};

\node (result) [text width=2.5cm, below of =solver, xshift=-1cm, yshift=0cm] {\textbf{Realizable/Unrealizable}};

\node (attracc) [process, text width=2.5cm, right of = solver,  xshift = 1cm] {Attractor acceleration};
\node (solveacc) [process, text width=2.5cm, below of =attracc, yshift=1cm] {Outer fixpoint acceleration};

\node (extract) [process, text width=2cm, left of = solver,  xshift = -.5cm] {Program extraction};
\node (absprog) [io, text width=1.4cm, left of = extract,  xshift = -.7cm] {Abstract reactive program};
\node (genc) [process, text width=2cm, below of =absprog, yshift=.6cm] {C program construction};

\draw [arrow] (issycomp) -- (llissyprob);
\draw [arrow] (llissyprob) -- (llissypars);
\draw [arrow] (llissypars) -- (spec);
\draw [arrow] ($(spec.north) - (2cm,0) $) -- ($(directconstr.south) - (.8cm,0) $);
\draw [arrow] ($(spec.north) + (0.13cm,0) $) -- (monconstr);
\draw [arrow] (monconstr) -- (monitor);
\draw [arrow] (directconstr) -- (formgame);
\draw [arrow] (monitor) -- (otfprod);
\draw [arrow] (formgame) -- (otfprod);
\draw [arrow] (otfprod) -- ($(enhgame.north) + (.65,0)$);
\draw [arrow] ($(enhgame.south) + (.65,0)$) -- (gameprod);
\draw [arrow] ($(spec.east) - (0.1,0.5)$) -- ($(gameprod.west) - (0,0.5)$);
\draw [arrow] (gameprod) -- (game);
\draw [arrow] (game) -- (gameinterface);

\draw [arrow] (rpgpars) -- (rpg);
\draw [arrow] (tslmttorpg) -- (rpg);
\draw [arrow] (rpg) -- (gameinterface);

\draw [arrow] (gameinterface) -- (solver);
\draw [doublearrow] (solver) -- (attracc);
\draw [doublearrow] (solver) -- (solveacc);

\draw [arrow] (solver) -- ($(result.north) + (1.35cm,0) $);

\draw [arrow] (solver) -- (extract);
\draw [arrow] (extract) -- (absprog);
\draw [arrow] (absprog) -- (genc);

\node (formulastogame) [draw=blue,very thick,rounded corners,fit = (directconstr) (monconstr) (formgame) (monitor) (otfprod), inner sep=3pt] {};
\begin{scope}[on background layer]
\node (symbgameconstruction) [draw=blue,very thick,rounded corners,fit = (formulastogame) (spec) (gameprod),inner sep=3pt,fill=blue!10] {};
\end{scope}

\begin{scope}[on background layer]
\node (synthesisengine) [draw=blue,very thick,rounded corners,fit = (solver) (extract) (attracc) (solveacc), inner sep=3pt,fill=blue!10] {};
\end{scope}

\node (issyf) [above of=issycomp,text width=.5cm,yshift=-.7cm] {\scalebox{1.9}{\faFile*[regular]}\newline \textbf{.issy}};
\draw [arrow] (issyf) -- (issycomp);

\node (llissyf) [left of=llissypars,text width=.5cm, xshift=.1cm] {\scalebox{1.9}{\faFile*[regular]}\newline \textbf{.llissy}};
\draw [arrow] (llissyf) -- (llissypars);

\node (rpgf) [left of=rpgpars,text width=.5cm, xshift=.0cm] {\scalebox{1.9}{\faFile*[regular]}\newline \textbf{.rpg}};
\draw [arrow] (rpgf) -- (rpgpars);

\node (tslmtf) [left of=tslmttorpg,text width=.5cm, xshift=.0cm] {\scalebox{1.9}{\faFile*[regular]}\newline \textbf{.tslmt}};
\draw [arrow] (tslmtf) -- (tslmttorpg);

\node (cf) [left of=genc,text width=.5cm, xshift=.3cm] {\scalebox{1.9}{\faFile*[regular]}\newline \textbf{.c}};
\draw [arrow] (genc) -- (cf);

\node [above of = otfprod,   xshift=-1.6cm, yshift=-.2cm] {\textcolor{blue}{\textbf{Translation to symbolic game} (Sec.~\ref{sec:to-game})}};

\node [above of = attracc,   yshift=-1.1cm,xshift=-1cm] {\textcolor{blue}{\textbf{Acceleration-based solver} (Sec.~\ref{sec:solver})}};
\end{tikzpicture}}

\end{center}
\caption{Architecture of \issy. Components depicted in blue and pink are part of the tool's implementation, those depicted in white are external.}
\label{fig:toolchain}
\end{figure}
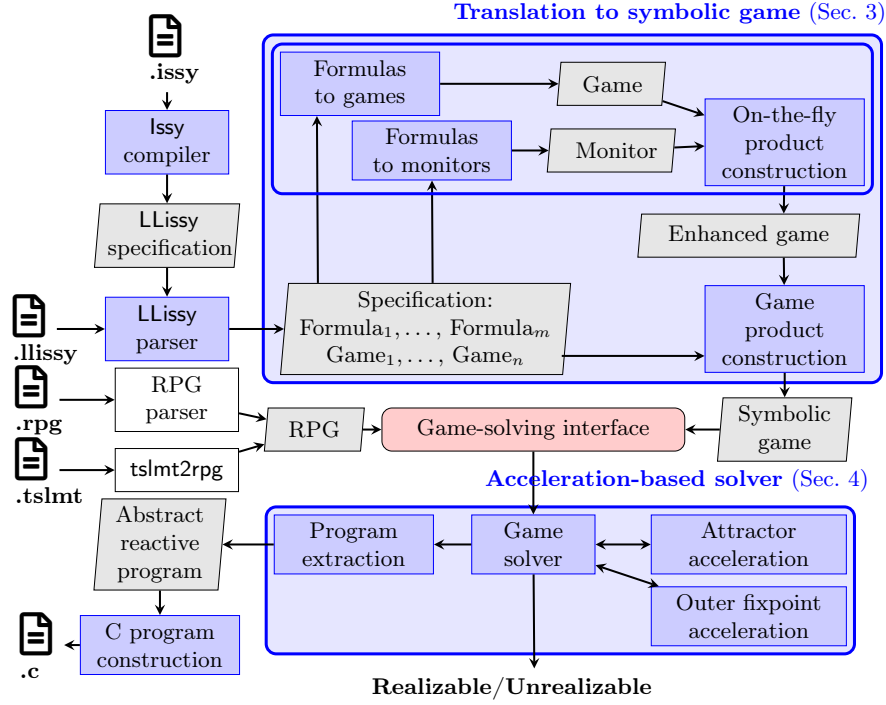

The architecture of \issy is shown in \Cref{fig:toolchain}.  We discussed the components translating a specification to a single synthesis game in \Cref{sec:to-game}.
Now we present the game solver underlying \issy, focusing on the novel  technical developments.

The approach behind the \issy solver builds on the method proposed in~\cite{HeimD24}.
The fist main difference to the prototype \rpgsolve from~\cite{HeimD24} is that \rpgsolve accepts RPGs,  a strictly more restricted class of symbolic games.  Furthermore,  the initial version of \rpgsolve does not support parity winning conditions.   
Our \Cref{ex:simple-example} cannot be modelled as an RPG,  because the system player has the power to select any real values as next-state values for the state variables.
Furthermore,  the specification in \Cref{ex:simple-example} translates to a parity game.
\issy's solver supports a more general symbolic game model, and also implements a symbolic method for infinite-state parity games based on fixpoint computation (a lifting of the classical Zielonka's algorithm~\cite{Zielonka98}).  Thus,  \issy is able to establish the realizability of the specification in Listing~\ref{lst:example} thanks to the new techniques it implements.

The crux to this is the acceleration technique introduced in~\cite{HeimD24}.
Naive fixpoint-based game-solving diverges on this example.
\emph{Attractor acceleration}~\cite{HeimD24} uses ranking arguments to establish that by iterating some strategy an unbounded number of times through some location, a player in the game can  enforce reaching a set of target states.  In \Cref{ex:simple-example},  attractor acceleration is used within the procedure for solving the parity game to establish that (under the respective constraints on the environment) from any state satisfying the formula \texttt{balanced},  a state where both \texttt{load1} and \texttt{load2} are in the bounded interval $[\frac{3}{10},\frac{9}{10}]$ can be enforced by the system player.  This argument is formalized as what is called an \emph{acceleration lemma}~\cite{HeimD24}.  From the interval $[\frac{3}{10},\frac{9}{10}]$,  the system player can then enforce reaching in a bounded number of steps a state where  \texttt{load1} and \texttt{load2} are zero.

We developed a novel method for generating acceleration lemmas and  implemented it in \issy in addition to that from~\cite{HeimD24}.
To search for acceleration lemmas,  \rpgsolve introduces uninterpreted predicates representing the lemmas' components, and collects SMT constraints asserting the applicability of the lemma.  Thus,  \rpgsolve would have to discover the formula \texttt{[load1 >= load2] \&\& [load1 <= 2 * load2]  ||[load2 >= load1] \&\& [load2 <= 2 * load1]} as\\part of the acceleration lemma, which it is not able to do within a reasonable timeout.
The alternative method implemented in \issy performs analysis of the game in order to generate candidate  acceleration lemmas.
First,  it analyzes the game in order to identify variables potentially making progress in a ranking argument.  For instance,  variables that remain unchanged in the relevant game locations can  be ruled out.  
Second, the new method uses the distance to the target set of states to generate ranking arguments for candidate acceleration lemmas.  Finally,  to search for a set of states where the  respective player can enforce the decrease of the distance, it uses symbolic iteration and SMT-based formula generalization.  As demonstrated for \Cref{ex:simple-example}, and more broadly by our experimental evaluation in \Cref{sec:experiments},  this new method for generating acceleration lemmas, which we call \emph{geometric acceleration}, is successful in many cases challenging for \rpgsolve. In \issy, geometric attractor acceleration is enabled by default, and the method can be switched using the parameter \texttt{\textcolor{blue}{-{}-accel-attr}}.

In addition to an alternative method for generating acceleration lemmas,  the \issy solver utilizes new techniques for their localization. Building on ideas in~\cite{SchmuckHDN24}, 
we restrict the size of the sub-games used for the acceleration lemma computation and project away variables that are not relevant in the respective subgame.  Unlike~\cite{SchmuckHDN24}, where this is done for pre-computing accelerations,  in \issy these localization techniques are applied on-the-fly during the main game solving.

\issy also provides support for strategy synthesis and extraction of C programs for realizable specifications.  The latter can be extended to other target languages,  utilizing the 
generic data structure for reactive program representation in \issy.

\issy\footnote{\url{https://github.com/phheim/issy}} is implemented in Haskell with focus on modularity and extensibility,  including detailed documentation.  Using the Haskell tool Stack,  building  \issy and getting its dependencies is seamless.
The external tools used are \texttt{Spot}~\cite{Duret-LutzRCRAS22}  for translation of LTL  to automata,  the $\mu$CLP solver \texttt{MuVal}~\cite{UnnoTGK23} and the Optimal CHC solver \texttt{OptPCSat}~\cite{GuTU23}  for the monitor construction,  and \texttt{z3}~\cite{Z3} for all SMT,  formula simplification, and quantifier elimination queries.

\section{Benchmarks and Evaluation}\label{sec:experiments}
\newcommand{\macroissycomp}{
\begin{tikzpicture}
\begin{axis}[%
    xmin=-2, xmax=1202, xmode=log, 
    ymin=-2, ymax=1202, ymode=log, 
    xlabel={Runtime (sec.) geom. accel. }, 
    ylabel={Runtime (sec.) unint. accel.  }, 
    width=\textwidth, 
    height=\textwidth%
    ]%
    \draw[gray] (axis cs:-2,-2) -- (axis cs:1202,1202);
    \addplot[only marks, mark=x, color=blue]  table [x=geom, y=unif, col sep=comma] {./data/plot-comp-accel.csv};
\end{axis}
\end{tikzpicture}}

\newcommand{\macrosweapcomp}{
\begin{tikzpicture}
\begin{axis}[%
    xmin=-2, xmax=1202, xmode=log, 
    ymin=-2, ymax=1202, ymode=log, 
    xlabel={Runtime (sec.) Issy (best)}, 
    ylabel={Runtime (sec.) sweap}, 
    width=\textwidth, 
    height=\textwidth%
    ]%
    \draw[gray] (axis cs:-2,-2) -- (axis cs:1202,1202);
    \addplot[only marks, mark=x, color=blue]  table [x=issy, y=sweap, col sep=comma] {./data/plot-comp-sweap.csv};
\end{axis}
\end{tikzpicture}}

\newcommand{\macrorpgplot}{
\begin{tikzpicture}
\begin{axis}[width=0.85\textwidth, height=3.6cm, ylabel near ticks, legend pos=outer north east, xmin=-5, xticklabel = {$\pgfmathprintnumber[precision=0]{\tick}s$}]
\addplot[., red] table[y=Cnt,x=Time,col sep=comma] {data/plot-rpgs-issy.csv}; 
\addplot[., blue] table[y=Cnt,x=Time,col sep=comma] {data/plot-rpgs-rpgsolve.csv}; 
\addplot[., brown] table[y=Cnt,x=Time,col sep=comma] {data/plot-rpgs-rpgstela.csv}; 
\addplot[., green] table[y=Cnt,x=Time,col sep=comma] {data/plot-rpgs-muval.csv}; 
\addplot[., gray] table[y=Cnt,x=Time,col sep=comma] {data/plot-rpgs-sweap.csv}; 
\addlegendentry{\scriptsize Issy (sel.)}
\addlegendentry{\scriptsize rpgsolve}
\addlegendentry{\scriptsize rpgstela}
\addlegendentry{\scriptsize muval}
\addlegendentry{\scriptsize (sweap)}
\end{axis}
\end{tikzpicture}}

\newcommand{\macrotslmtplot}{
\begin{tikzpicture}
\begin{axis}[width=0.85\textwidth, height=3.6cm, ylabel near ticks, legend pos=outer north east, xmin=-5,  xticklabel = {$\pgfmathprintnumber[precision=0]{\tick}s$}]
\addplot[., red] table[y=Cnt,x=Time,col sep=comma] {data/plot-tslmt-issy.csv}; 
\addplot[., blue] table[y=Cnt,x=Time,col sep=comma] {data/plot-tslmt-tslmt2rpg.csv}; 
\addplot[., gray] table[y=Cnt,x=Time,col sep=comma] {data/plot-tslmt-sweap.csv}; 
\addplot[., brown] table[y=Cnt,x=Time,col sep=comma] {data/plot-tslmt-raboniel.csv}; 
\addlegendentry{\scriptsize Issy (sel.)}
\addlegendentry{\scriptsize tslmt2rpg}
\addlegendentry{\scriptsize (sweap)}
\addlegendentry{\scriptsize Raboniel}
\end{axis}
\end{tikzpicture}}

\newcommand{\macrosummarytable}{
\begin{tabular}{|l||r|r|r||r|} \hline
Benchmark set      & \Cref{figure:rpg-compare} & \Cref{figure:tslmt-compare} & New \issy{} &  hard in \cite{HeimD24} 
\\\hline\hline
Total number of benchmarks          &  105 & 94 &  55 &   8 \\\hline\hline
Geometric accel.                    &  \textbf{88} &  40 & 40 & 4 \\\hline 
Uninterpreted-predicate accel.      &  62 &  32 & 40 & 2 \\\hline 
Geometric accel. + monitor pruning  &   - &  \textbf{72} & \textbf{42} & \textbf{5} \\\hline
Uninterpreted-predicate accel.      &   - &   - & 41 & - \\
~+ monitor pruning                  &     &     &    &   \\ \hline 
\end{tabular}
}

\begin{figure}[t!]
\begin{minipage}[t]{0.47\textwidth}
\centering
\scalebox{0.92}{\macrorpgplot}
\captionof{figure}{\footnotesize Solved instances of 105 RPGs from \cite{HeimD24,SchmuckHDN24,AzzopardiPSS24} within given time (in sec.). 4 instances were only solved by each \issy{} and sweap, and 1 instance was only solved by \muval.}\label{figure:rpg-compare}
\end{minipage}
\hfill
\begin{minipage}[t]{0.47\textwidth}
\centering
\scalebox{0.92}{\macrotslmtplot}

\captionof{figure}{\footnotesize Solved instances of 94 TSL-MT from \cite{MaderbacherB22,AzzopardiPSS24,HeimD25} within given time (in sec.). 13 instances were only solved by sweap, 2 instances were only solved by \issy{}, and 1 instance was only solved by \raboniel{}.}\label{figure:tslmt-compare}
\end{minipage}
\end{figure}

We evaluated \issy{} experimentally,  comparing to  
\raboniel{}\footnote{\url{https://doi.org/10.5281/zenodo.7602503}},
sweap\footnote{\url{https://github.com/shaunazzopardi/sweap}, commit: 1275a759},
\muval{}\footnote{\url{https://github.com/hiroshi-unno/coar}, commit: dc094f04},
\rpgsolve{}\footnote{\url{https://doi.org/10.5281/zenodo.10939871}} and
\tslmtrpg{}\footnote{\url{https://doi.org/10.5281/zenodo.13939202}},
thus covering all types of techniques. 
We did not compare to \temos{} as past experiments~\cite{HeimD24,HeimD25} show that it is outperformed by \raboniel{}.
The other tools are either not available, unable to build, or do not accept input files.
Also, we did not compare to~\cite{KatisFGGBGW18} since it is restricted to safety specifications, and in~\cite{SamuelDK21} is mostly outperformed by \gensys.
Neither did we compare to~\cite{MaderbacherWB24} as it focuses on efficient synthesis from GR(1), not on handling unbounded behaviour, and the implementation was not available to us.

For \issy{} we use four configurations: with the novel geometric acceleration or the existing acceleration with uninterpreted predicates,  and with or without monitor-based simplification (\texttt{\textcolor{blue}{-{}-pruning}} 2 or 0, resp.) when applied to specifications with formulas.
The later is because the effectiveness of pruning varies~\cite{HeimD25}.

We used an extensive set of benchmarks\footnote{\url{https://github.com/phheim/infinite-state-reactive-synthesis-benchmarks}} (contributions welcome!) containing the RPG benchmarks from~\cite{HeimD24,SchmuckHDN24} and the TSL-MT benchmarks~\cite{MaderbacherB22,HeimD25}, some of which can not be solved by existing tools.
Furthermore, we included the benchmarks created by the authors of sweap~\cite{AzzopardiPSS24} (in their format) as well as their manually encoded versions in the RPG and TSL-MT formats.
We also created 50 new benchmarks in the new \issy{} format which combine formuals and games and can only be used by \issy{}.

We partitioned the set of benchmarks according to the type of specifications (games or temporal formulas) the tools are applicable to  according to \Cref{table:compare-input}. We apply \muval{} on RPGs via an automatic encoding of the games as fixpoint equations which is similar to the one in~\cite{UnnoSTK20} and modular w.r.t. the game locations, i.e.\ it uses one sub-equation per location.
All experiments were run on AMD EPYC processors,  with one core, 4GB of memory, and 20 minutes wall-clock-time for each benchmarking run.

\begin{figure}[t!]
\begin{minipage}[t]{0.46\textwidth}
\centering
\scalebox{0.95}{\macrosweapcomp}
\captionof{figure}{Comparison of sweap and \issy on the sweap benchmarks~\cite{AzzopardiPSS24}, manually encoded in~\cite{AzzopardiPSS24} as RPGs or TSL-MT. 23 instances were only solved by \issy and 19 instances were only solved by sweap.}\label{figure:issy-vs-sweap}
\end{minipage}
\hfill
\begin{minipage}[t]{0.46\textwidth}
\centering
\centering
\scalebox{0.95}{\macroissycomp}
\captionof{figure}{Comparison of the existing and new attractor acceleration in \issy on all RPG, TSL-MT, and new \issy benchmarks. 57 instances were only solve by the new attractor acceleration and 5 instances were only solved by the existing one.}\label{figure:accel-compare}
\end{minipage}
\end{figure}

\begin{table}[t!]\footnotesize
\begin{center}
\macrosummarytable
\end{center}
\caption{Benchmark instances solved by the four different \texttt{Issy} configurations.}\label{table:summary-data}
\end{table}

\Cref{figure:rpg-compare} and \Cref{figure:tslmt-compare} show the comparisons on 105 RPG and 94 TSL-MT specifications, respectively.
For \issy{} we show the best time for \emph{checking realizability} for each benchmark across the four different configurations.  As shown in \Cref{table:summary-data},  the best time for \issy is usually with geometric acceleration.
We ran sweap only on the benchmarks to which it is applicable and are available in its own format, which uses a different formalism.
Therefore,  we show additionally in  \Cref{figure:issy-vs-sweap} the comparison to sweap only on the set of those 148 benchmarks. 
We note that sweap is performing synthesis, while the results for \issy are for checking realizability, as we could not let sweap only check for realizability.
The evaluation results demonstrate that \issy{} mostly outperforms the existing prototypes,  and has matured well beyond the prototypes it stems from.

In addition to \Cref{table:summary-data}, \Cref{figure:accel-compare} provides a detailed comparison between the new geometric and the existing uninterpreted-predicate-based acceleration methods (without pruning) on all benchmarks. 
It shows that geometric acceleration is effective, without making the existing  acceleration method obsolete.
We also ran \issy in synthesis mode (\texttt{\textcolor{blue}{-{}-synt}}) with geometric acceleration, especially as synthesis for uninterpreted-predicate-based acceleration is known to be hard~\cite{HeimD24}. 
Out of the 130 benchmarks that geometric accleration  determined to be realizable,  \issy could synthesize C-programs for 106 of them within the given resource bounds.
This difference stems from the fact that \issy does heavy simplifications and might need to synthesize Skolem functions.

In summary,  the results show \issy's competitiveness with the state of the art. \issy's comprehensive framework, together with the public collection of benchmarks provide a basis for further development of techniques and tools.

\begin{credits}

\subsubsection{Data Availability Statement.}
The software generated during and analysed during the current study is available in the Zenodo repository~\url{https://doi.org/10.5281/zenodo.15308725}.

\subsubsection{\discintname}
The authors have no competing interests to declare that are relevant to the content of this article.
\end{credits}

\bibliographystyle{splncs04}
\bibliography{main.bib}

\newpage

\appendix

\section{Full Description of the \issy Format}\label{sec:format-issy-full}
\subsubsection{\issy  specification}

\grammarindent2.2cm

An \issy specification consists of variable declarations, formula specifications, game specifications, and macro definitions for better readability.

\begin{grammar}

<spec> ::= (<vardecl> | <logicspec> | <gamespec> | <macro>)*
 
\end{grammar}

\subsubsection{Variable Declarations}

Variable declarations determine the variables that can be used in the formula and game specifications. A variable declaration specifies whether the variable is input controlled by the environment, or is a state variable controlled by the system. The currently supported data types are bool, int and real. Variables have global scope and can be used in any of the formula and game specifications in the file.

\begin{grammar}

<vardecl> ::= (`input' | `state') <type> <identifier> 

 <type>    ::= `int' | `bool' | `real'

\end{grammar}

\subsubsection{Formula Specifications}

A formula specification is a list of RPLTL formulas (as defined in~\cite{HeimD25}), prefixed by the keywords \textsf{assume} and \textsf{assert}, denoting constraints on the environment and the system respectively. They use temporal operators (next \textsf{X}, eventually \textsf{F}, globally \textsf{G}, until \textsf{U}, weak until \textsf{W}, release \textsf{R}) like LTL, but with quantifier-free first-order atoms  instead of Boolean propositions. The precedence is like in TLSF~\cite{JacobsPS23}. A formula specification is an implication with antecedent the conjunction of the assumptions and consequent the conjunction of the asserts.

\begin{grammar}

<logicspec> ::=  `formula' `{' <logicstm>$^*$  `}'  

 <logicstm> ::= (`assert' | `assume') <rpltl>

<rpltl>    ::=  <atom> \alt `(' <rpltl> `)' \alt <uopt> <rpltl> \alt <rpltl> <bopt> <rpltl>

<uopt>    ::= `!' | `F' | `X' | `G'

<bopt>    ::= `&&' | `||' | `->' | `<->' | `U' | `W' | `R'

\end{grammar}

\subsubsection{Game Specifications}

A game specification defines the type of winning condition and the initial location of the defined game. Locations (which can be thought of as the values of a local program counter) have names, colors (used in the winning condition), and domain terms (which constrain the possible values of the variables). Transition definitions determine the possible transitions between locations. They are labeled by formulas, which specify under what conditions a transition can be taken and what is its effect. 
Location names have local scope -- if a location name appears in multiple games, those are unrelated. The domain formula associated with each location acts like an invariant restricting the set of possible variable valuations in states with this location.

\begin{grammar}

<gamespec>  ::= `game' <wincond> `from' <identifier> `{' ( <locdef> | <transdef>)*`}' 

<wincond>  ::= `Safety' | `Reachability' | `Buechi'  | `CoBuechi' | `ParityMaxOdd'

<locdef>   ::= `loc' <identifier> [<nat>] [`with' <formula>]

<transdef>  ::= `from'  <identifier> `to' <identifier> `with' <formula>

\medskip

<formula>    ::=  <atom>  \alt '(' <formula> ')' \alt <uop> <formula> \alt <formula> <bop> <formula> 

<uop>      ::=  `!' 

<bop>      ::=  `&&' | `||' | `->' | `<->'

\end{grammar}
\noindent
The precedence of the logical operators is as follows (from high to low): 
$$\textsf{ \{!\} > \{\&\&\} > \{||\} > \{-> (ra)\} > \{<-> (ra)\}}$$

The winning condition is defined via the locations' colors, which are natural numbers. For Safety, Reachability, Buechi, and CoBuechi a location is respectively safe, target, Buechi accepting, coBuechi accepting (should be visited eventually always) if and only if the number is greater than zero. For ParitMaxOdd the number is the color in the parity game.

An \issy specification can contain multiple logical and game specifications, which are interpreted conjunctively. 
At most one of them is allowed to be a non-safety game, respectively not a syntactic safety RPLTL formula.

\subsubsection{Atomic Predicates}

Atomic predicates are defined as follows.

\begin{grammar}

<atom>   ::=  <apred> \alt <bconst> \alt  <identifier>['''] \alt `havoc' `('<identifier>* `)' \alt `keep' `(' 
<identifier>* `)'

<bconst>  ::= `true'  | `false'

<apred>   ::= `[' <pred> `]'

<pred>    ::= <const> \alt <identifier>['''] \alt `(' <pred> `)' \alt <auop> <pred> \alt <pred> <abop> <pred>

<const>   ::= <nat>   | <rat>

<auop>    ::= `*' | `+' | `-' | `/' | `mod' | `=' | `<' | `>'| `<=' | `>='

<abop>   ::= `-' | `abs'

\end{grammar}

\noindent
The precedence of the operators is as follows (from high to low): 
$$\textsf{\{abs\} > \{*, /, mod\} > \{+, -\} > \{<, >, =, <=, >=\}}$$

\subsubsection{Macros}

Macros make writing specifications more convenient.

\begin{grammar}
<macro>  ::= `def' <identifier> `=' <formula> | <apred>
\end{grammar}

\noindent
Macros can be used in all $\langle \mathit{rpltl}\rangle$,  $\langle \mathit{formula}\rangle$, and $\langle\mathit{pred}\rangle$. However, for usage in $\langle\mathit{pred}\rangle$ the marco term has to be a single predicate term.

\subsubsection{Identifiers and Numerical Constants}

Identifiers and numerical constants are defined as follows.

\begin{grammar}
<identifier>      ::= <alpha> (<alpha> | <digit> | `_')*

<nat>             ::= <digit>+

<rat>            ::= <digit>+ '.' <digit>+
\end{grammar}

\subsubsection{Comments}

\issy specifications support C-like single-line (\textsf{/ /})and multi-line comments (\textsf{/ *}).
Note that comments cannot be nested.

\noindent

\newpage

\section{The LLissy Format}\label{sec:format-llissy}

In order to be easy to parse, readable with reasonable effort, and to be similar to the SMTLib-format, \llissy uses s-expressions.

Only single line comments exist which start with \textsf{';'} and span to the end of the line. 
Newlines are \textsf{'\textbackslash r\textbackslash n', '\textbackslash n \textbackslash r', ' \textbackslash r'} and \textsf{'\textbackslash n'}. 
However, when generating \llissy automatically \textsf{'\textbackslash n'} should be used. 
Similarly \textsf{' '} (Space) and \textsf{'\textbackslash t'} (Tabs) are both non-newline white-spaces. 
However, only \textsf{' '} should be used upon generation. 
The following productions define identifiers and natural numbers.
\begin{grammar}
<ALPHA> ::= `a'...`z' | `A'...`Z' 

<DIGIT> ::= `0'...`9'

<ID>    ::= <ALPHA> (<ALPHA> | <DIGIT> | `_')*

<PID>   ::= <ID> ['$\sim$']

<NAT>   ::=  <DIGIT>+

<RAT>   ::=  <DIGIT>+ `.' <DIGIT>+
\end{grammar}

\noindent
Note that all of these should be parsed greedily until a white-space, '(', ')', or the end-of-file occurs.

A \llissy specification consists of lists of variable declarations, formula specifications and game specifications. The variables declarations include all variables used in all games and formulas. 
The formula and game specifications are interpreted conjunctively. 
However, at most one game or formula can be a non-safety game or non-safety formula.

\begin{grammar}
<SPEC> ::= `(' `(' <VARDEC>* `)' `(' <FSPEC>* `)' `(' <GSPEC>*  `)' `)'
\end{grammar}

\noindent
A variable declaration declares an input or state variable and its respective type
\begin{grammar}
<VARDEC>  ::= `(' `input' <TYPE> <ID> `)' | `(' `state' <TYPE> <ID> `)'

<TYPE>   ::=  `Int' | `Bool' | `Real'
\end{grammar}

A formula specification is a pair of assumption and guarantee lists. Each element is an RP-LTL formula.
The assumptions come first, and each of the two lists is interpreted as a conjunction. 
\begin{grammar}
<FSPEC>   ::= `(' `(' <FORMULA>* `)' `(' <FORMULA>* `)' `)'

<FORMULA> ::= `(' `ap' <TERM>`')' \alt `(' <UOP> <FORMULA> `)' \alt `(' <BOP> <FORMULA> <FORMULA> `)' \alt (<NOP> <FORMULA>*)

<UOP>     ::= `X' | `F' | `G' | `not'

<BOP>     ::= `U' | `W' | `R'

<NOP>     ::= '`and' | `or'
\end{grammar}

A game specification consists of a list of location definitions, transition definitions from one location to another location, and an objective definition.
The objective defines the initial location and the winning condition. Each location is annotated with a natural number. For Safety, Reachability, Buechi, and CoBuechi a location is safe, target, Buechi accepting, coBuechi accepting iff the number is greater than zero. 
For ParitMaxOdd the number is the color in the parity game.

\begin{grammar}
<GSPEC>    ::= `(' `(' <LOCDEF>* `)' `(' <TRANSDEF>* `)' <OBJ> `)'

<LOCDEF>   ::= `(' <ID> <NAT> <TERM> `)'

<TRANSDEF> ::= `(' <ID> <ID> <TERM> `)'

<OBJ>      ::= `(' <ID> (`Safety' | `Reachability ' | `Buechi' | `CoBuechi' \newline \phantom{aaaaaaaaaaaaaaaa} | `ParityMaxOdd') `)'
\end{grammar}

A term is basically like in the SMT-Lib-2 format without quantifiers, lambda, and let expressions. Similar rules for typing apply.
Only variables declared initially are allowed to be free variables, and additionally primed version (with $\sim$) of the state variables.

\begin{grammar}
<TERM>   ::= `(' <OP> <TERM>* `)' | <PID> | <CONSTS>

<OP>     ::= `and ' | `or' | `not' | `ite' | `distinct' | `=>' |
         `=' | `<' | `>'| `<=' | `>=' |
         `+' | `-' | `*' | `/' | `mod' | `abs' | `to_real' 

<CONSTS> ::= <RAT> | <NAT> | `true' | `false'
\end{grammar}

\end{document}